# Crystallization behaviors in superionic conductor Na$_3$PS$_4$

Hiroshi Nakajima[1], Hirofumi Tsukasaki[1], Jiong Ding[1], Takuya Kimura[2], Takumi Nakano[2], Akira Nasu[2], Akihiko Hirata[3, 4], Atsushi Sakuda[2], Akitoshi Hayashi[2], and Shigeo Mori[1]

*[1]Department of Materials Science, Osaka Prefecture University, Sakai, Osaka 599-8531, Japan*

*[2]Department of Applied Chemistry, Osaka Prefecture University, Sakai, Osaka 599-8531, Japan*

*[3]Department of Materials Science, Waseda University, Shinjuku, Tokyo 169-8555, Japan*

*[4]Kagami Memorial Research Institute for Materials Science and Technology, Waseda University, Shinjuku, Tokyo 169-0051, Japan*

Abstract

All-solid-state batteries using sodium are promising candidates for next-generation rechargeable batteries due to the limited lithium resources. A practical sodium battery requires an electrolyte with high conductivity. Cubic Na$_3$PS$_4$ exhibiting high conductivity of over $10^{-4}$ S cm$^{-1}$ is obtained by crystallizing amorphous Na$_3$PS$_4$ synthesized by ball milling. Amorphous Na$_3$PS$_4$ crystallizes in a cubic structure and then is transformed into a tetragonal phase upon heating. In this study, in situ observation by transmission electron microscopy demonstrates that the crystallite size drastically increases during the transition from the cubic phase to the tetragonal phase. Moreover, an electron diffraction analysis reveals that amorphous domains and nano-sized crystallites coexist in the cubic Na$_3$PS$_4$ specimen, while the tetragonal phase contains micro-sized crystallites. The nano-sized crystallites and the composite formed by crystallites and amorphous domains are most likely responsible for the increase in conductivity in the cubic Na$_3$PS$_4$ specimens.

# 1. Introduction

All-solid-state batteries are promising as rechargeable batteries due to their higher energy densities and increased safety compared with conventional lithium-ion batteries using organic liquid electrolytes. Since an all-solid-state battery requires a solid electrolyte with high ionic conductivity, many solid electrolytes have been synthesized with the aim of improving the battery performance. Recent studies have shown that sulfide-based lithium-ion conductors such as $Li_3PS_4$, $Li_7P_3S_{11}$, and $Li_{10}GeP_2S_{12}$ exhibit high conductivity (above $10^{-3}$ S cm$^{-1}$), even higher than that of organic liquid electrolytes in some cases [1–5]. However, the growing demand for portable devices brings another concern for battery development: the lithium resource shortage. Thus, lithium-free solid electrolytes are highly desirable to fabricate all-solid-state batteries.

In 2012, Hayashi *et al.* reported that cubic $Na_3PS_4$ precipitated from an amorphous phase shows high conductivity of over $10^{-4}$ S cm$^{-1}$ [6]. They also demonstrated that cold pressing can densify the electrolyte and a rechargeable cell with $Na_3PS_4$ has excellent cycle performance at room temperature. Besides the obvious advantage of abundant sodium resources, $Na_3PS_4$ is a promising electrolyte because its conductivity and stability against humid atmosphere can be increased by substitution, reaching conductivity values ranging from 10 to 40 mS cm$^{-1}$ [7,8].

Amorphous $Na_3PS_4$ prepared by ball milling is known to crystallize in the cubic structure (space group $I\bar{4}3m$) at approximately 180°C and in the tetragonal structure ($P\bar{4}2_1c$) at 350°C [6,9–11]. However, several crystallographic characteristics remain elusive, such as the microstructures and real-space distribution of the crystallites precipitated from the amorphous phase. Besides, heat-treatment conditions change the peak widths of X-ray diffraction and nuclear magnetic resonance [12–14]. The measurements suggest the presence of amorphous fractions that affect the conductivity, as demonstrated in a lithium sulfide electrolyte [15]. Previous studies have emphasized the need to directly observe the crystallite size and the amorphous fraction. Furthermore, to the best of our knowledge, direct observation of the crystallization process in sodium superionic conductors, which would provide useful information for the synthesis of sodium electrolytes, has not been performed. In this study, we performed in situ heating observation of the crystallization process and microstructures in $Na_3PS_4$ by transmission electron microscopy (TEM). The results reveal the presence of a mixture of amorphous domains and crystallites with nano-sized diameters in the cubic $Na_3PS_4$ specimens. Furthermore, grain coarsening occurs during the transition from the cubic phase to the tetragonal phase. The difference of conductivity between the cubic and tetragonal phases can be explained in terms of the observed properties.

## 2. Materials and Methods

To observe the microstructure of $Na_3PS_4$, amorphous $Na_3PS_4$ was heated in an electric furnace (annealed specimens) or in the microscope (in situ observation). Amorphous $Na_3PS_4$ was prepared by a mechanochemical process using a planetary ball mill as previously described [6]. Annealed specimens were synthesized by heating amorphous $Na_3PS_4$ at 240°C or 480°C in a dry Ar atmosphere to the cubic or tetragonal structures, respectively. The ionic conductivities of $Na_3PS_4$ specimens were determined using an impedance analyzer (SI-1260, Solartron). The prepared amorphous and annealed $Na_3PS_4$ powders were uni-axial pressed at 360 MPa to form pellets for impedance measurements.

Transmission electron microscopy observation was conducted using a field-emission transmission electron microscope operated at 200 kV (JEM-2100F, JEOL Co. Ltd.). The images and diffraction patterns were recorded with a charge-coupled device camera (Gatan, US1000). To prepare the samples for observation, powder specimens were crushed and dispersed on a carbon grid supported by copper mesh. The grids were attached to a double-tilt vacuum TEM holder (Gatan model 648). These procedures were done in a glove box filled with dry Ar gas. The in situ observation was conducted with a double-tilt heating holder (Gatan). The temperature was increased from 20°C to 400°C in steps of 30°C–60°C, and each temperature was maintained for 30 min to reach thermal equilibrium. Crystallites were segmented using machine learning to calculate the average crystallite sizes [16]. Nano-beam electron diffraction (NBED) was performed with a beam diameter of approximately 10 nm. The NBED mapping was obtained using the scanning function of the microscope in annular dark-field mode. The diffraction patterns were recorded using imaging plates (Ditabis, Micron) with high-dynamic and large-angle ranges to perform reduced pair distribution function (PDF) analysis. The diffraction patterns for the PDF analysis were obtained from the polycrystalline regions of crushed specimens without using a direct beam stopper. The reduced PDF was obtained using the electron diffraction software [17].

The degree of crystallinity was determined from the electron diffraction patterns [18]. The crystallinity $\chi(T)$ at a temperature $T$ can be defined as follows:

$$\chi(T) = \frac{I_{crystal}(T)}{I_{crsytal}(T) + I_{amorphous}(T)} \quad (1)$$

where $I_{crystal}(T)$ and $I_{amorphous}(T)$ represent the integral intensities of the Bragg peaks from the crystallites and the halo rings from the amorphous domains, respectively. Since the crystalline fraction is not directly proportional to the Bragg reflection intensity, the crystallinity derived from this equation defines the crystallization behaviors resulting from the temperature changes.

## 3. Results and discussion

We performed in situ TEM observation to investigate the crystallization behavior of amorphous $Na_3PS_4$ upon heating. Figures 1 and 2 show the changes in the TEM images and the corresponding diffraction patterns with increasing the temperature, respectively. The observation of a halo-ring pattern in Fig. 2 at 20°C confirmed the amorphous nature of the $Na_3PS_4$ structure before heating. During heating from 20°C to 100°C, no structural change was observed and the amorphous structure was maintained. However, nano-crystallites appeared at approximately 110°C, which can be observed as bright areas in the dark-field image of Fig. 1 (indicated by yellow arrows). Accordingly, the electron diffraction pattern at 110°C shows Bragg reflections due to the crystallization process. The crystallization proceeded upon further heating, and Debye-ring patterns appeared from 180°C to 340°C. Within this temperature range, the nano-size of the crystallites was maintained. Besides, the presence of the halo-ring patterns suggests that crystallites coexisted with amorphous domains. Noticeably, the crystallite size drastically increased at 340°C, as evidenced by the large grains appearing as bright areas (marked by red dot lines) in the corresponding dark-field image. This temperature corresponds to the onset of the transition from the cubic phase to the tetragonal phase, as revealed by a differential thermal analysis [6]. Finally, a large grain of more than 500 nm appeared at 400°C. A comparison of the dark-field images at 400°C and 110°C–270°C revealed the presence of several nano-crystallites in the area where the large grain appeared, which suggests that this large grain was formed by merging several nano-crystallites into a single grain. This is also observed in Supplementary Fig. 1, which shows other grains grown in other areas. These images prove the formation of large grains from numerous nano-crystallites at approximately 400°C. Meanwhile, the external shape was found to remain unchanged from 20°C to 400°C, demonstrating that the secondary particle remained intact, while the primary particles transformed their shape and size upon heat treatment. The electron diffraction patterns confirmed this crystallization behavior. The Bragg reflection spots observed at 400°C changed into discrete peaks with stronger intensity, which is in accordance with the transformation of the nano-crystallites into large grains. This crystallization behavior was reproduced in another amorphous particle (Supplementary Fig. 2). These behaviors demonstrate that crystallites maintained their nano-size in the cubic phase, whereas they merged into large grains in the tetragonal phase.

To gain more insight into these changes, we examined the size and crystallinity of the crystallites (Fig. 3). The inset in Fig. 3(a) shows that the crystallite size, which ranged from 10 to 30 nm in the cubic phase, increased linearly with increasing the temperature from 110°C to 340°C. Upon further heating from 340°C to 400°C, a rapid increase in the crystallite size was observed, as can be extracted from the dark-field images shown in Fig. 1, reaching an average grain diameter of 260 nm at 400°C. The temperature dependence of the crystallinity exhibited a similar behavior [Fig. 3(b)]. The total crystalline area increased with increasing temperature as demonstrated in Supplementary Fig. 3, similar to the crystallinity. Thus, the crystallinity increased slightly upon heating in the cubic phase from 110°C to 340°C. As suggested by the diffraction patterns in this temperature range (Fig. 2), the cubic phase showed lower crystallinity (60%) than the tetragonal phase due to the coexistence of nano-crystallites and amorphous domains. In the tetragonal phase, by contrast, almost all areas crystallize and large grains were detected, as shown in the dark-field image at 400°C (Fig. 1). Thus, the crystallinity in the tetragonal phase was approximately 100%.

The in situ observation suggests that the cubic $Na_3PS_4$ specimen exhibited both nano-crystallites and amorphous domains. Furthermore, we performed NBED on the specimen annealed at 240°C. Owing to the nano-sized beam, NBED allows obtaining diffraction patterns from only a single crystallite or an amorphous domain, which enables the separation of different crystallographic phases [19]. Figure 4(a) displays an annular dark-field scanning image of $Na_3PS_4$ annealed at 240°C. The marked region was mapped by NBED as shown in Fig. 4(b). The areas 1, 4, 5, 8, and 9 show Bragg reflections, demonstrating the presence of crystallites. Conversely, the halo-ring patterns observed in the areas 2, 3, 6, and 7 are indicative of an amorphous structure. Thus, the cubic $Na_3PS_4$ specimen contained both cubic crystallites and amorphous domains.

To confirm the crystallization behavior observed in the in situ heating experiment, we investigated specimens annealed at 240°C and 480°C. The bright-field images of Fig. 5(a) and 5(d) show the meandering shape of the specimens annealed at 240°C and sharp clear edges for those annealed at 480°C, respectively. This difference between the secondary particles results from the primary particles. The diffraction pattern of specimens annealed at 240°C exhibits Debye–Scherrer rings, showing that the primary particles comprised nano-crystallites with various crystal orientations [Fig. 5(b)]. The dark-field image shown in Fig. 5(c) shows nano-crystallites with an average size of approximately 15 nm. This size is consistent with that observed in the in situ experiment, as plotted in Fig. 3(a). Conversely, the diffraction patterns of Fig. 5(e) and (f) show single-domain patterns of the $[\bar{1}\bar{3}2]$ and $[01\bar{2}]$ orientations, respectively. As shown in Supplementary Fig. 4, diffraction patterns characteristic of single-domain areas were also observed for other grains. These results demonstrate that the specimens annealed at 480°C contained large grains in the tetragonal phase, which also agrees with the in situ observation. The average crystal diameter was 1160 nm, which matches the trend of grain coarsening extracted from the in situ observation results [Fig. 3(a)]. The crystallite sizes of cubic and tetragonal $Na_3PS_4$ were comparable to those estimated by X-ray analysis [9,13]. The crystallinity of specimens annealed at 480°C was almost 100% due to the single-crystal patterns of the large grains. Therefore, the specimens annealed at 240°C and 480°C reproduced the results of the in situ observation.

Next, we confirmed the local structure of $Na_3PS_4$ by PDF analysis in each phase. Figure 6(a) shows the profiles of the electron diffraction patterns of amorphous and annealed specimens in each phase. The amorphous specimen gave rise to a low-intensity halo-ring pattern. Conversely, the specimens annealed at 240°C and 480°C produced Bragg reflections. The peak positions in Fig. 6(b) agree with the simulations based on the cubic ($I\bar{4}3m$) and tetragonal ($P\bar{4}2_1c$) structures. Figure 6(c) exhibits the reduced PDF $G(r)$ in each phase. In the reduced PDF, a peak position represents the bond length between two atoms, and the peak area is related to the coordination number. Note that since electron diffraction is subjected to multiple scattering and inelastic scatterings, the peak intensity and the density of the specimen cannot be evaluated quantitatively in this study [20,21]. Hence, we focused only on the peak positions. In the amorphous phase, the first and third peaks around 2.0 and 3.3 Å correspond to the P–S and S–S bond lengths in the $PS_4$ tetrahedron of the cubic structure, respectively, which confirms the presence of the short-range order of $PS_4$ in the amorphous phase. Besides, the 2.85 Å peak in the amorphous phase corresponds to the Na–S bond. A remarkable damping of peaks at distances greater than 3.8 Å demonstrates the lack of long-range order in the amorphous phase. However, broad peaks around 3.8, 5.5, and 6.7 Å with the same positions as those of the peaks of the specimens annealed at 240°C and 480°C can still be observed. Conversely, peaks indicating long-range order appeared above 4 Å in the specimens annealed at 240°C and 480°C, and the peak positions of both specimens were the same. Therefore, the local structure remained unchanged during the transition from the cubic phase to the tetragonal phase, which is consistent with previous X-ray PDF studies [9,22]. Notably, the previous results of PDF analyses were obtained for voluminous bulk specimens. Conversely, in the present electron diffraction study, PDF data were obtained from areas with a diameter below 1 μm, which confirmed that a local structure of the cubic phase is similar to that of the tetragonal phase even when obtaining the PDF data from a nano-scale region.

The reasons for the higher conductivity of the cubic phase compared with that of the tetragonal phase can be discussed as follows: Compared with tetragonal $Na_3PS_4$ prepared by the solid-state reaction method, cubic $Na_3PS_4$ specimens fabricated by ball milling showed significantly higher conductivity. Recent studies suggest that the high conductivity stems from the formation of sodium vacancies in off-stoichiometry specimens caused by the ball milling process [22–25]. Hence, tetragonal and cubic $Na_3PS_4$ synthesized by ball milling should show similar conductivity because they possess the same density of sodium vacancies. Besides, the similar conductivity of cubic and tetragonal $Na_3PS_4$ structures was predicted by a density functional theory study [23]. However, some studies have reported that cubic $Na_3PS_4$ shows slightly higher conductivity than tetragonal $Na_3PS_4$ and that the annealing conditions affect the conductivity [12,13]. In this study, the conductivity of the cubic $Na_3PS_4$ specimen showed a room-temperature conductivity of $1.3 \times 10^{-4}$ S cm$^{-1}$, which was higher than that of tetragonal $Na_3PS_4$ ($1.9 \times 10^{-5}$ S cm$^{-1}$). According to the X-ray analysis [13], these results originate from small crystallites and a mixture of crystallites and amorphous domains in the cubic $Na_3PS_4$ specimen. Our real-space observation results demonstrate that the crystallite size ranged from 10 to 30 nm and the size change due to the temperature increase was small in the cubic phase. Conversely, the size rapidly increased in the tetragonal phase. These differences could explain the conductivity differences between each phase because nano-sized crystallites can increase the conductivity [26,27]. Note that the crystallinity of the cubic $Na_3PS_4$ specimen is over 60% in the annealed specimen as shown in Fig. 3, and thus the cubic $Na_3PS_4$ phase should be interconnected based on the percolation theory [28,29].

The grain coarsening and almost complete crystallization observed in $Na_3PS_4$ is a strikingly different behavior compared with sulfide-based lithium-ion electrolytes such as $Li_3PS_4$ and $Li_7P_3S_{11}$. The latter exhibit amorphous fractions and nano-sized crystallites even after heating the specimens to high temperatures above the crystallization temperatures [15,30,31]. In these materials, the crystalline phases with high conductivity are metastable, and crystallites precipitate from the amorphous phases. Similarly, cubic $Na_3PS_4$ prepared by ball milling is metastable, and the amorphous phase remains in the cubic phase. However, $Na_3PS_4$ can be fully crystallized upon reaching the stable tetragonal phase. The coexistence of amorphous domains and crystallites with nano-meter sizes might be a common characteristic in metastable phases precipitated from amorphous electrolytes.

Furthermore, we demonstrated in real space that the cubic $Na_3PS_4$ specimen contains an amorphous phase, which could contribute to the enhancement of the conductivity. The amorphous phase itself cannot increase the conductivity because the amorphous $Na_3PS_4$ prepared by ball-milling in this study has a lower conductivity of $4.1 \times 10^{-6}$ S cm$^{-1}$ than cubic and tetragonal $Na_3PS_4$ specimens. However, a solid electrolyte with a mixture of amorphous domains and crystallites was reported to exhibit higher conductivity than separate crystalline or amorphous electrolytes [32]. Ionic movements are shown to be faster at the amorphous–crystalline interfaces, contributing to the conductivity enhancement. Moreover, the presence of amorphous domains decreases the grain boundary resistance because the contacts within the electrolyte improve, leading to an increase in conductivity [13,33]. Our microscopy results illustrate that the cubic $Na_3PS_4$ specimen is composed of a homogeneous mixture of cubic crystallites and amorphous domains, whereas the tetragonal phase comprises single domains. Therefore, cubic $Na_3PS_4$ specimens should have better performance than tetragonal $Na_3PS_4$ ones owing to the presence of amorphous domains. The nano-sized cubic $Na_3PS_4$ phase, which is interconnected in amorphous domains, has a high $Na^+$ ion conductivity. A fast $Na^+$ ion migration at amorphous–crystalline interfaces is another possibility for explaining the experimental results of high conductivity in the cubic $Na_3PS_4$ specimen. A detailed understanding of the ionic motion at the amorphous-crystal interfaces requires further kinetic-property measurements and structural analysis of cubic $Na_3PS_4$ specimens.

## 4. Conclusions

The cubic $Na_3PS_4$ electrolyte synthesized by ball milling is composed of nano-crystallites with a diameter ranging from 10 to 30 nm, and its crystallite size and crystallinity slightly increase upon heating. Moreover, crystallites and amorphous domains coexist in the cubic phase. The presence of nano-sized crystallites without grain coarsening seems to be a common characteristic of metastable phases precipitated from amorphous electrolytes prepared by ball milling. Conversely, the tetragonal phase comprises crystals with single domains. The small crystallite size and the presence of amorphous domains contribute to the enhancement of the conductivity in the cubic Na$_3$PS$_4$ electrolyte. This study sheds some light on the understanding of Na$_3$PS$_4$ microstructures, which paves the way for the synthesis of solid electrolytes with high conductivity.

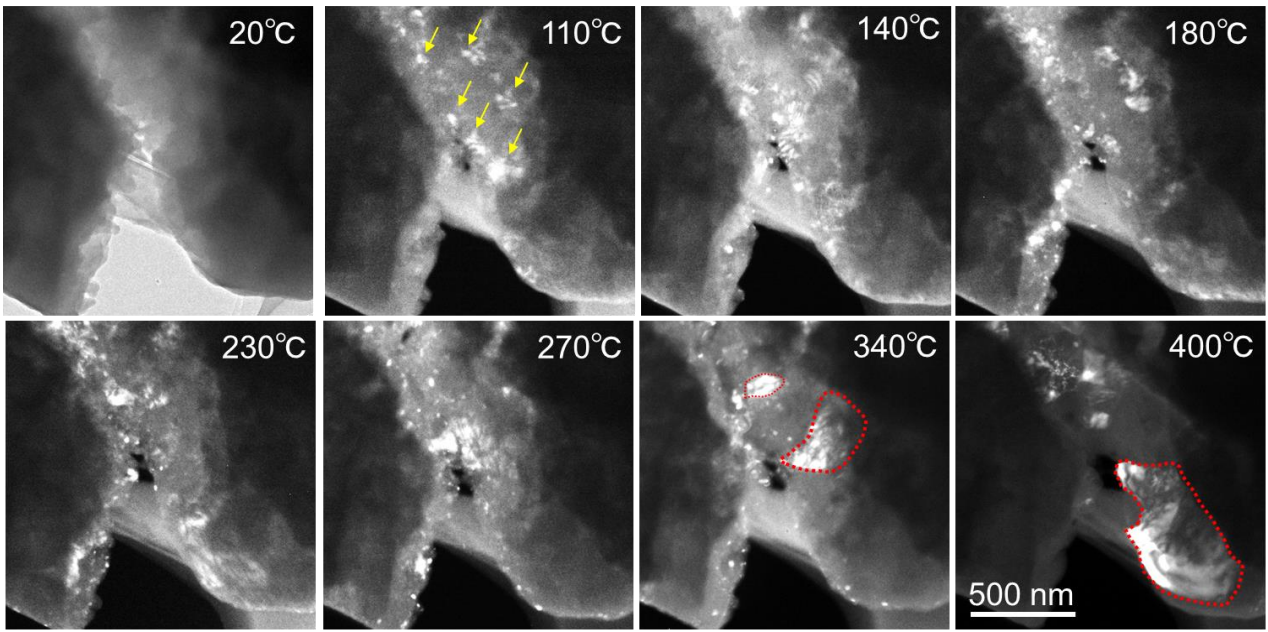

Fig. 1. In situ transmission electron microscopy observation of the crystallization behavior of amorphous Na$_3$PS$_4$ with increasing the temperature. The images at 20°C and 110°C–400°C were obtained using bright- and dark-field techniques, respectively. The yellow arrows mark crystallites at 20°C. The red dot lines indicate large grains that appeared upon heating.

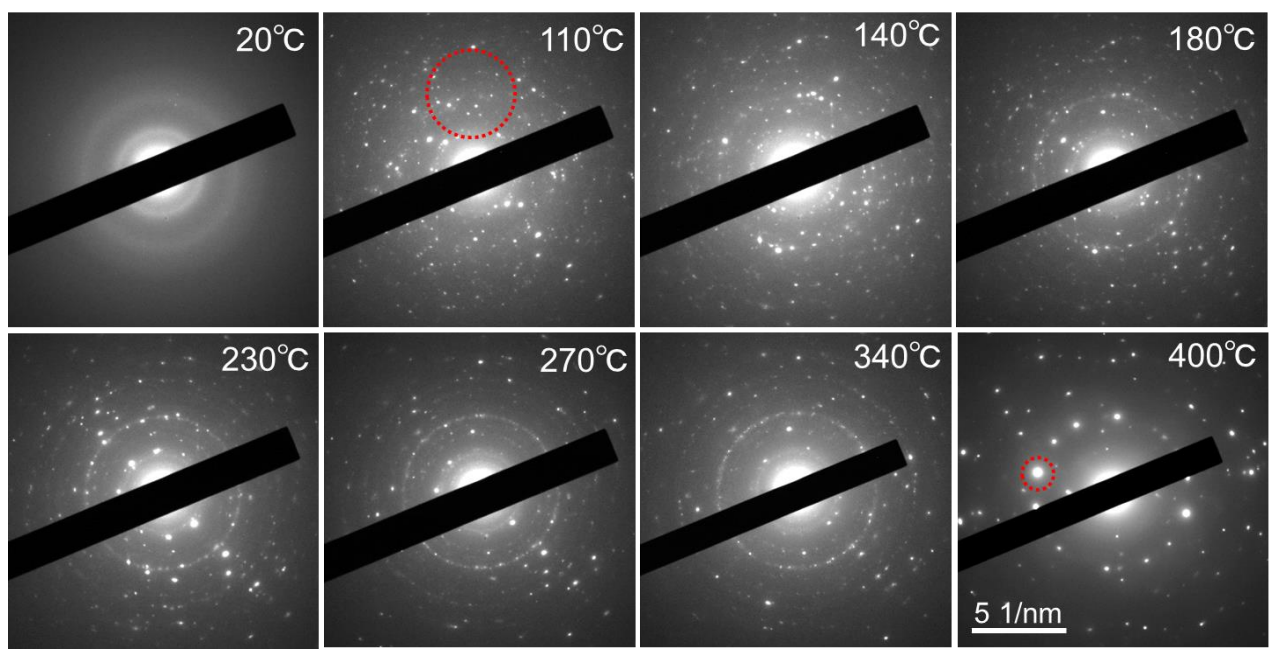

Fig. 2. Electron diffraction patterns upon increasing the temperature in amorphous $Na_3PS_4$. The patterns were obtained from the areas corresponding to Fig. 1. The black rectangle is a direct beam stopper. The reciprocal space ranges used for the dark-field images of Fig. 1 are marked with red circles. The same reciprocal space range was used for the dark-field images at temperatures ranging from 110°C to 340°C.

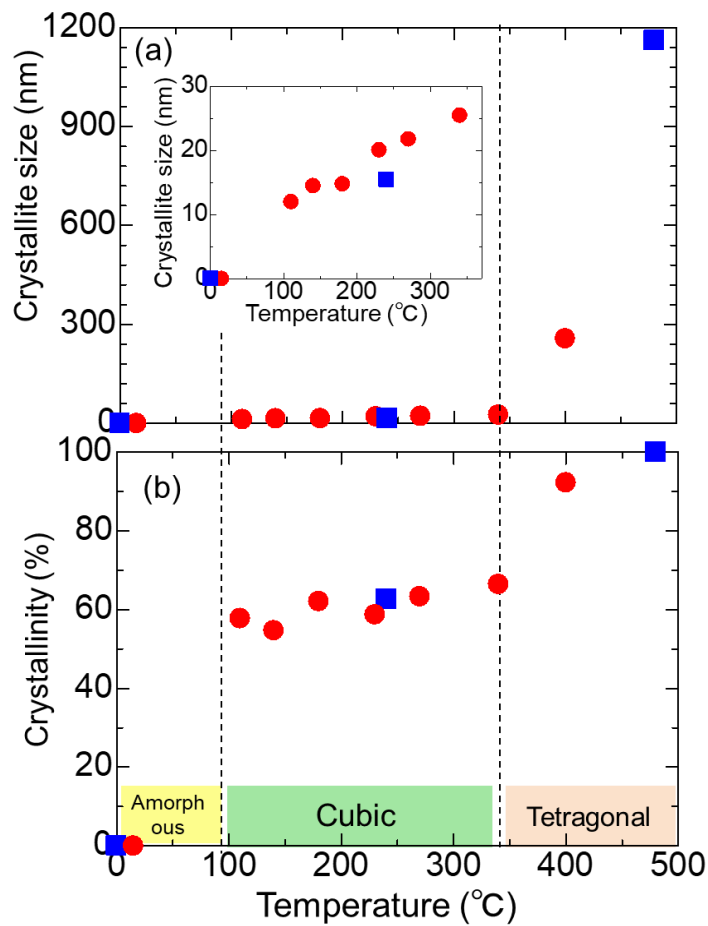

Fig. 3. Temperature dependence of (a) crystallite size and (b) crystallinity. The crystallite size and crystallinity were calculated from Figs. 1 and 2, respectively. The circles and squares represent the values measured from in situ data of Fig. 1 and annealed specimens of Fig. 5, respectively. The inset of (a) shows a magnified graph from 0°C to 370°C.

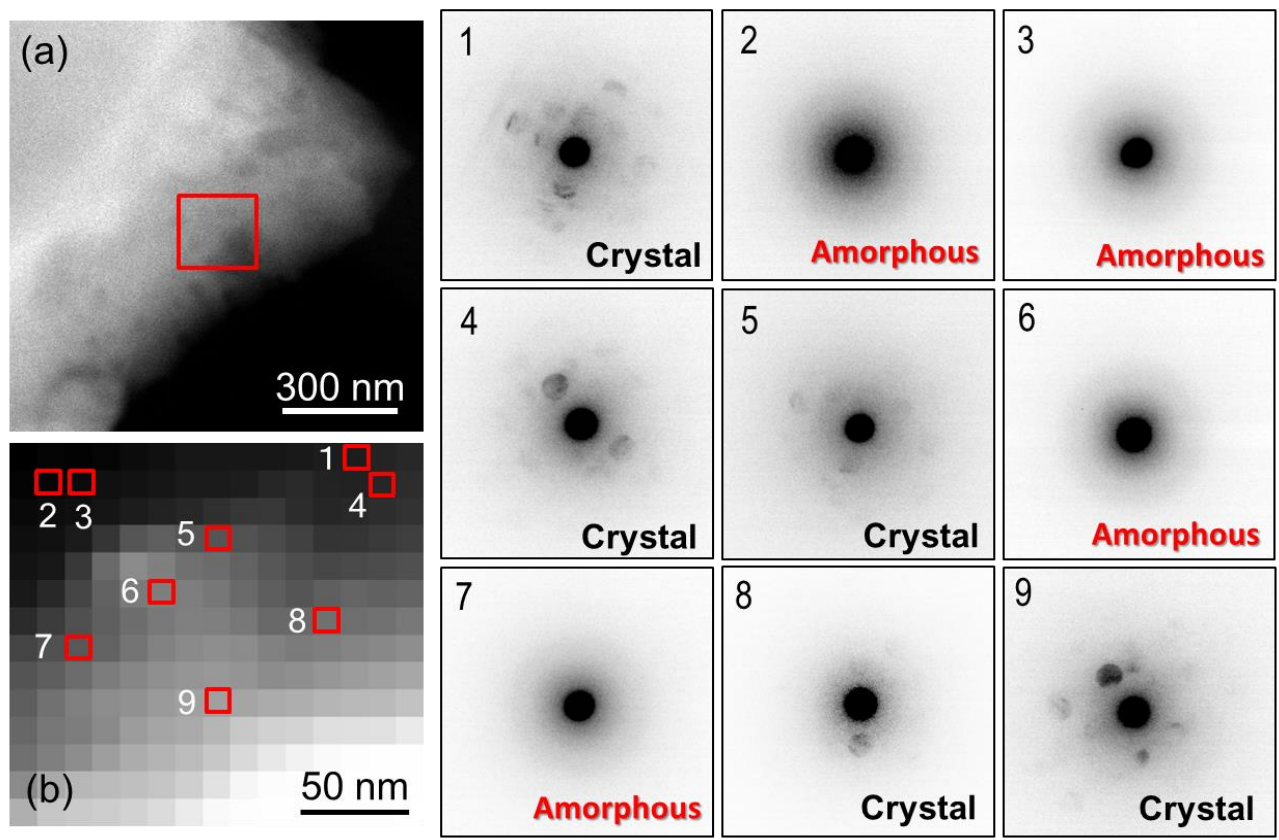

Fig. 4. (a) Annular dark-field scanning image of Na$_3$PS$_4$ annealed at 240°C. (b) Nano-beam electron diffraction intensity mapping from the squared region of (a). The right panels are nano-beam diffraction patterns corresponding to the marked regions. The observation was conducted at room temperature. The probe size was approximately 10 nm.

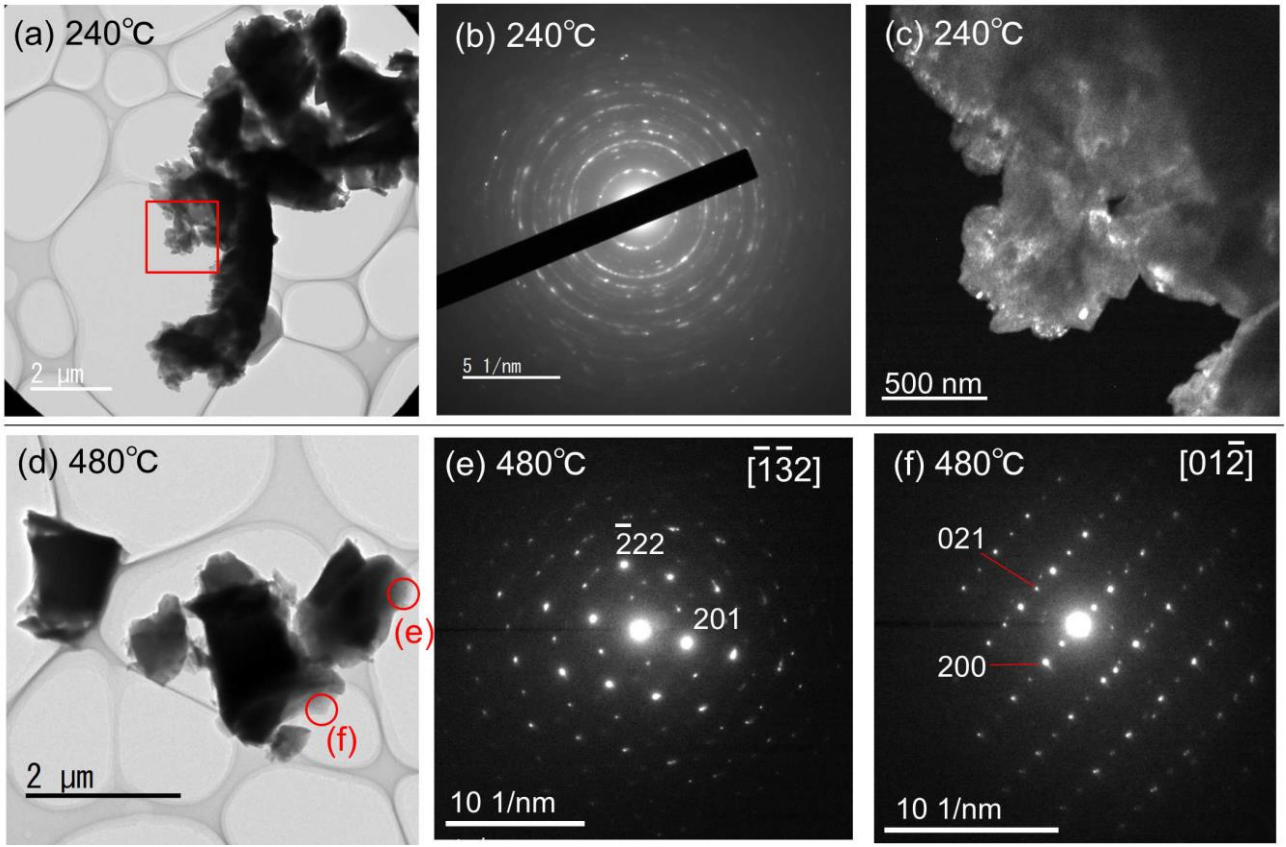

Fig. 5. Ex situ transmission electron microscopy images of specimens annealed at (a)–(c) 240°C and (d)–(f) 480°C. (a) and (d) Bright-field images. (b), (e) and (f) Electron diffraction patterns. (c) Dark-field image. In (d), the red marks indicate the regions from which the diffraction patterns were obtained. The observation was performed at room temperature.

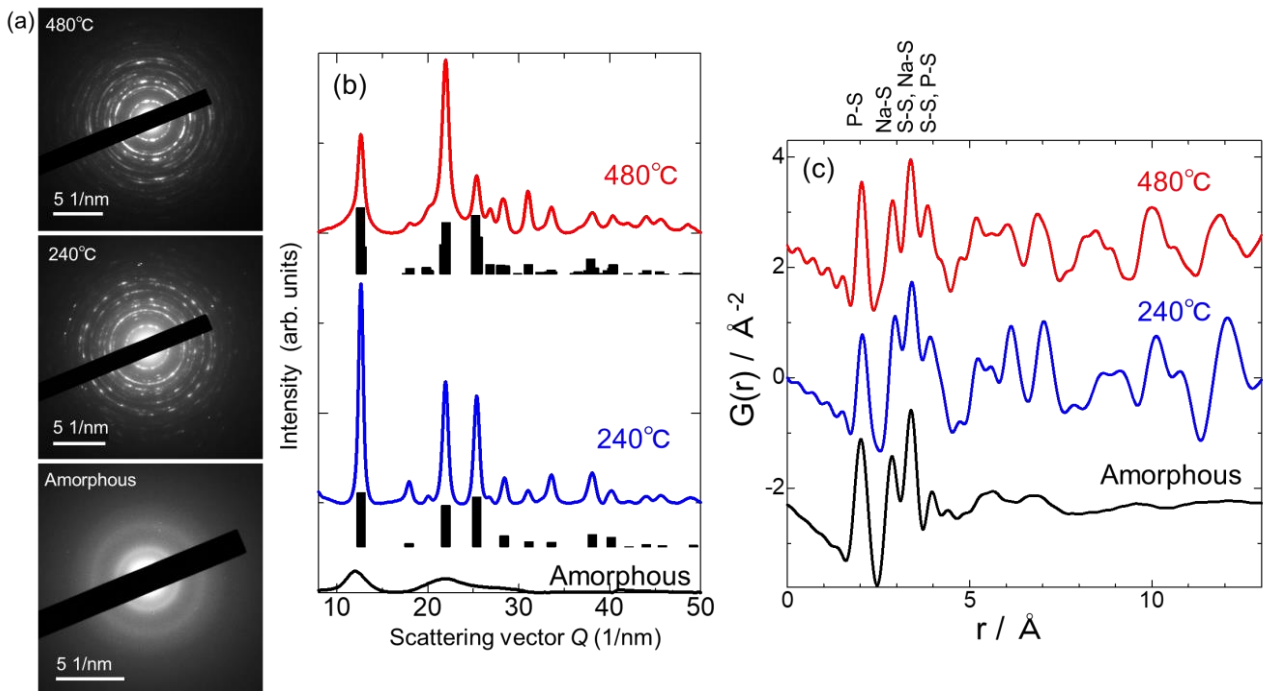

Fig. 6. Comparison of the electron diffraction patterns of amorphous and annealed specimens. (a) Electron diffraction patterns. (b) Radial averaged intensity profiles of the diffraction patterns (a). (c) Reduced pair distribution function in each phase.